\documentclass[conference]{IEEEtran}
\IEEEoverridecommandlockouts
\usepackage{cite}
\usepackage{amsmath,amssymb,amsfonts}
\usepackage{algorithm}
\usepackage{graphicx}
\usepackage{textcomp}
\usepackage{xcolor}
\usepackage{amsmath,amssymb,amsfonts}

\usepackage{mathtools}
\usepackage{algorithm}
\usepackage{hyperref}
\hypersetup{
    colorlinks=true,
    linkcolor=red,
    filecolor=magenta,      
    urlcolor=blue,
    citecolor=violet
}

\usepackage{xcolor, soul}

\usepackage[noend]{algpseudocode} 
\usepackage{graphicx}
\usepackage{tabularx}
\usepackage{multirow}

\usepackage{amsthm}
\usepackage{url}
\usepackage{bm}
\usepackage{tikz} 
\usepackage{wrapfig}
\usepackage{textcomp}
\usepackage{xcolor}
\usepackage{caption}
\usepackage{subcaption}
\usepackage{comment}
\usepackage{mathtools}
\usepackage{xparse}
\usepackage{arydshln}
\usepackage{amssymb}
\usepackage{pifont}

\usepackage{tablefootnote}

\usepackage{color,soul}
\usepackage{lipsum}

\NewDocumentCommand{\vect}{ O{} O{} m }{\bm{#3}\ifthenelse{\isempty{#1}}{}{^{(#1)}}\ifthenelse{\isempty{#2}}{}{_{#2}}}

\NewDocumentCommand{\mat}{ O{} O{} m }{\bm{#3}\ifthenelse{\isempty{#1}}{}{^{(#1)}}\ifthenelse{\isempty{#2}}{}{_{#2}}}

\NewDocumentCommand{\ten}{ O{} O{} m }{\bm{\mathcal{#3}}\ifthenelse{\isempty{#1}}{}{^{(#1)}}\ifthenelse{\isempty{#2}}{}{_{#2}}}
\begin{document}

\title{Distributed Out-of-Memory SVD\\on CPU/GPU Architectures\\
}

\author{\IEEEauthorblockN{Ismael Boureima}
\IEEEauthorblockA{\textit{Theoretical Division}\\
\textit{LANL}\\
Los Alamos, U.S\\
iboureima@lanl.gov}
\\
\IEEEauthorblockN{Nick Solovyev}
\IEEEauthorblockA{\textit{Theoretical Division}\\
\textit{LANL}\\
Los Alamos, U.S\\
nks@lanl.gov}
\and
\IEEEauthorblockN{Manish Bhattarai}
\IEEEauthorblockA{\textit{Theoretical Division}\\
\textit{LANL}\\
Los Alamos, U.S \\
ceodspspectrum@lanl.gov}
\\
\IEEEauthorblockN{Hristo Djidjev}
\IEEEauthorblockA{\textit{Information Systems}\\
\textit{LANL}
Los Alamos, U.S \\
and IICT, Sofia, Bulgaria\\
djidjev@lanl.gov}
\and
\IEEEauthorblockN{Maksim E. Eren}
\IEEEauthorblockA{\textit{Theoretical Division}\\
\textit{LANL}\\
Los Alamos, U.S\\
maksim@lanl.gov}
\\
\IEEEauthorblockN{Boian S. Alexandrov}
\IEEEauthorblockA{\textit{Theoretical Division}\\
\textit{LANL}\\
Los Alamos, U.S\\
boian@lanl.gov}

}

\maketitle

\begin{abstract}
We propose an efficient, distributed, out-of-memory implementation of the truncated singular value decomposition (t-SVD) for heterogeneous (CPU+GPU) high performance computing (HPC) systems.
Various implementations of SVD have been proposed, with most only estimate the singular values as the estimation of the singular vectors can significantly increase the time and memory complexity of the algorithm. 
In this work, we propose an implementation of SVD based on the power method, which is a truncated singular values and singular vectors estimation method. Memory utilization bottlenecks in the power method used to decompose a matrix $\mat{A}$ are typically associated with the computation of the Gram matrix $\mat{A}^T\mat{A}$ , which can be significant when $\mat{A}$ is large and dense, or when $\mat{A}$ is super-large and sparse.
The proposed implementation is optimized for out-of-memory problems where the memory required to factorize a given matrix is greater than the available GPU memory. We reduce the memory complexity of $\mat{A}^T\mat{A}$ by using a batching strategy where the intermediate factors are computed block by block, and 
we hide I/O latency associated with both host-to-device (H2D) and device-to-host (D2H) batch copies by overlapping each batch copy with compute using CUDA streams. Furthermore, we use optimized \textit{NCCL} based communicators to reduce the latency associated with collective communications (both intra-node and inter-node). In addition, sparse and dense matrix multiplications are significantly accelerated with GPU cores (or tensors cores when available), resulting in an implementation with good scaling. We demonstrate the scalability of our distributed out of core SVD algorithm to successfully decompose dense matrix  of size 1TB and sparse matrix of size 128~PB with 1e-6 sparsity.
\end{abstract}


\section{Introduction}

\textit{Singular value decomposition} (SVD) decomposes a matrix $\mat{A}$ of size $m\times n$ into two orthogonal matrices $\mat{U}$ of size $m\times m$, $\mat{V}$ of size $n\times n$, and a diagonal matrix $\Sigma$ of size $m\times n$, which is a non-negative matrix, such that $\mat{A}=\mat{U}\Sigma\mat{V}^T$. The diagonal elements of $\Sigma$ are also called \textit{singular values}, whereas $\mat{U}$ and $\mat{V}$ contain left and right \textit{singular vectors} of $\mat{A}$, respectively. The power method is one way of estimating the SVD. The power method computes the first $k$  monotonically decreasing singular values and the corresponding vectors, saving the computational need to estimate the complete singular values and vectors\cite{bentbib2015block}. The power method based SVD is also known as \textit{truncated SVD} \cite{hansen1990truncated} is given as $\mat{A}=\mat{U}_k\Sigma_k\mat{V}_k^T$ where $\Sigma_k$ is a diagonal matrix comprising singular values $\sigma_1>\sigma_2>....>\sigma_k$ and $\mat{U}_k$ and $\mat{V}_k$ are the orthogonal matrices of sizes $m\times k$ and $k\times n$, respectively, whose columns are the left and right singular vectors corresponding to the first $k$ largest singular values. The ability to estimate the first $k$ large singular values and corresponding singular vectors addresses the challenge in estimating all the singular values and vectors. The pseudocode of the truncated SVD algorithm is shown in Algorithms~\ref{alg:svd_trunc} and~\ref{alg:svd_1d}.

\begin{algorithm}
  \scriptsize
	\caption{\small SVD($\mat{A},\epsilon,k=-1$) - Truncated SVD} 
	\label{alg:svd_trunc}
	\textbf{Require: $\mat{A}$} $\in \mathbb{R}_{+}^{m \times n}$ and a scalar $\epsilon$ where $k$ is an optional parameter.
	\begin{algorithmic}[1]
	\State $m,n$ = shape($\mat{A}$)
	\If{k==-1}
	\State $k$ = $\operatorname{min}(m,n)$
	\EndIf
	 	\State  $\mat{U},\mat{V},\mat{\sigma}$ = [],[],[] \Comment{Initialize as empty arrays}
    \For{$l$ in $[1,k]$}
    \State $\mat{X}$=$\mat{A}$
       \If {$l>1$}
       \State  $\mat{X}$ = $\mat{X} - \mat{U}[:l]diag(\mat{\sigma}[:l])\mat{V}[:l]^T$
       \EndIf
    \If{$m>n$}
    \State $\mat[l][]{V}$= $\operatorname{SVD\_1D}{(\mat{X},\epsilon)}$
    \State $\mat[l][]{U}$ = $\mat{A}@\mat[l][]{V}$ \Comment{@ stands for matrix multiplication operation}
    \State $\mat[l][]{\sigma}$ = $||\mat[l][]{U}||$
    \State  $\mat[l][]{U}$ = $\frac{\mat[l][]{U}}{\mat[l][]{\sigma}}$
    \Else{}
     \State $\mat[l][]{U}$= $\operatorname{SVD\_1D}{(\mat{X},\epsilon)}$
    \State $\mat[l][]{V}$ = $\mat{A}^T@\mat[l][]{U}$
    \State $\mat[l][]{\sigma}$ = $||\mat[l][]{V}||$
    \State  $\mat[l][]{V}$ = $\frac{\mat[l][]{V}}{\mat[l][]{\sigma}}$
    \EndIf
\EndFor
	\end{algorithmic} 
	\textbf{Ensure:} $\mat{U} \in \mathbb{R}^{m \times k}$,$\mat{\sigma} \in \mathbb{R}^{k}$, $\mat{V} \in \mathbb{R}^{n \times k}$\\
	\textbf{Ensure:} $\mat{U}\mat{U}^T = \mat[][m\times m]{I}$,$\mat{V}\mat{V}^T = \mat[][n\times n]{I}$ where $\mat{I}$ is Identity matrix and $diag(\mat{\Sigma})$ = $\mat{\sigma}$ where $\mat{\sigma} = \{\sigma^1,\sigma^2,...\sigma^k\}$ and $\mat{\Sigma} \in \mathbb{R}^{k \times k}$ 

\end{algorithm}

\begin{algorithm}
  \scriptsize
    \caption{\small SVD\_1D($\mat{X}$,$\epsilon$)}
    \label{alg:svd_1d}
    \textbf{Require: $\mat{X}$} $\in \mathbb{R}_{+}^{m \times n}$ and a scalar $\epsilon$.
	\begin{algorithmic}[1]
	\State $m,n$ = shape($\mat{X}$)
	\State  $k$ = $\operatorname{min}(m,n)$
	\State $\mat{x}\approx N(\mat{0},\mat{1})$ where $\mat{x},\mat{0},\mat{1} \in \mathbb{R}_{+}^{k}$ \Comment{Sample $x$ from a multivariate normal distribution of mean $\mat{0}$ and variance $\mat{1}$}
     \State $\mat{x}$ = $\frac{\mat{x}}{||x||}$ \Comment{Normalize $x$}
	\State set $\mat[][0]{v}=\mat{x}$ 
	\If{$m>n$}
	\State $\mat{B}$=$\mat{X}^T@\mat{X}$ \Comment{$@$ is matrix multiplication operator}
	\Else{}
	\State $\mat{B}$=$\mat{X}@\mat{X}^T$
	\EndIf
	\While {true}

	\State $\mat[][1]{v}$ = $\mat{B}@\mat[][0]{v}$
	\State $\mat[][1]{v}$ = $\frac{\mat[][1]{v}}{||\mat[][1]{v}||}$ \Comment{$||.||$ represents $l_2$ norm}
	\If {$|\mat[][0]{v}@\mat[][1]{v}| \ge 1-\epsilon$} \Comment{$|.|$ is modulus operator}
	 \State  return $\mat[][1]{v}$
	\EndIf
		\State $\mat[][0]{v}$=$\mat[][1]{v}$
	\EndWhile
    	\end{algorithmic} 
\end{algorithm}

SVD provides an advantage in decomposing large-scale datasets into low-dimensional factors, thus allowing compressed representations. As computing power has increased with the introduction of modern GPUs and TPUs, SVD (along with many other ML frameworks) has seen significant acceleration on this hardware. However, the ever-growing volume of data produced by social networks, medical applications, and experimental simulations has led to SVD decomposition bottlenecks in single computing hardware as the data does not fit in memory. In addition to these memory constraints, decomposition of such large-scale datasets requires significant computational resources along with data storage and movement. Almost every work in the relevant literature aims to solve only one or two of these challenges where the results do not show good scalability. This is due to the significant communication cost associated with data movement across different computing elements, which often exceeds the computation cost of these algorithms.

This mandates the need for distributed algorithms with the additional ability for out-of-memory computation that would enable performing computation over distributed hardware efficiently without memory bottleneck while also computing the singular vectors. In such a design, the data blocks not used in computation reside on the external disk or outside the memory of each node's modern GPU/TPU hardware. Furthermore, to make an efficient design, there is a need to overlap the data IO with computation operation to minimize the overall latency in performing decomposition. In this work, we devise a novel parallel SVD framework called pyDSVD-GPU, which combines batching for out-of-memory and tiling for distributed computation of large-sized sparse/dense datasets while reducing the communication and data movement cost on CPU/GPU heterogeneous hardware. The main contributions of this paper include:
\begin{itemize}
    \item The first novel distributed algorithm with out-of-memory support for SVD for sparse and dense matrices on CPU/GPU hardware.
    \item The first NCCL Communicator accelerated SVD decomposition tool in distributed GPUs.
    \item Custom SVD algorithm for decomposition of extra large, sparse datasets of sizes up to 128~PB and dense datasets of size up to 1TB. 
\end{itemize}

\begin{figure*}[ht!]
\centering
    \begin{minipage}[b]{0.5\linewidth}
        \includegraphics[width=.8\linewidth]{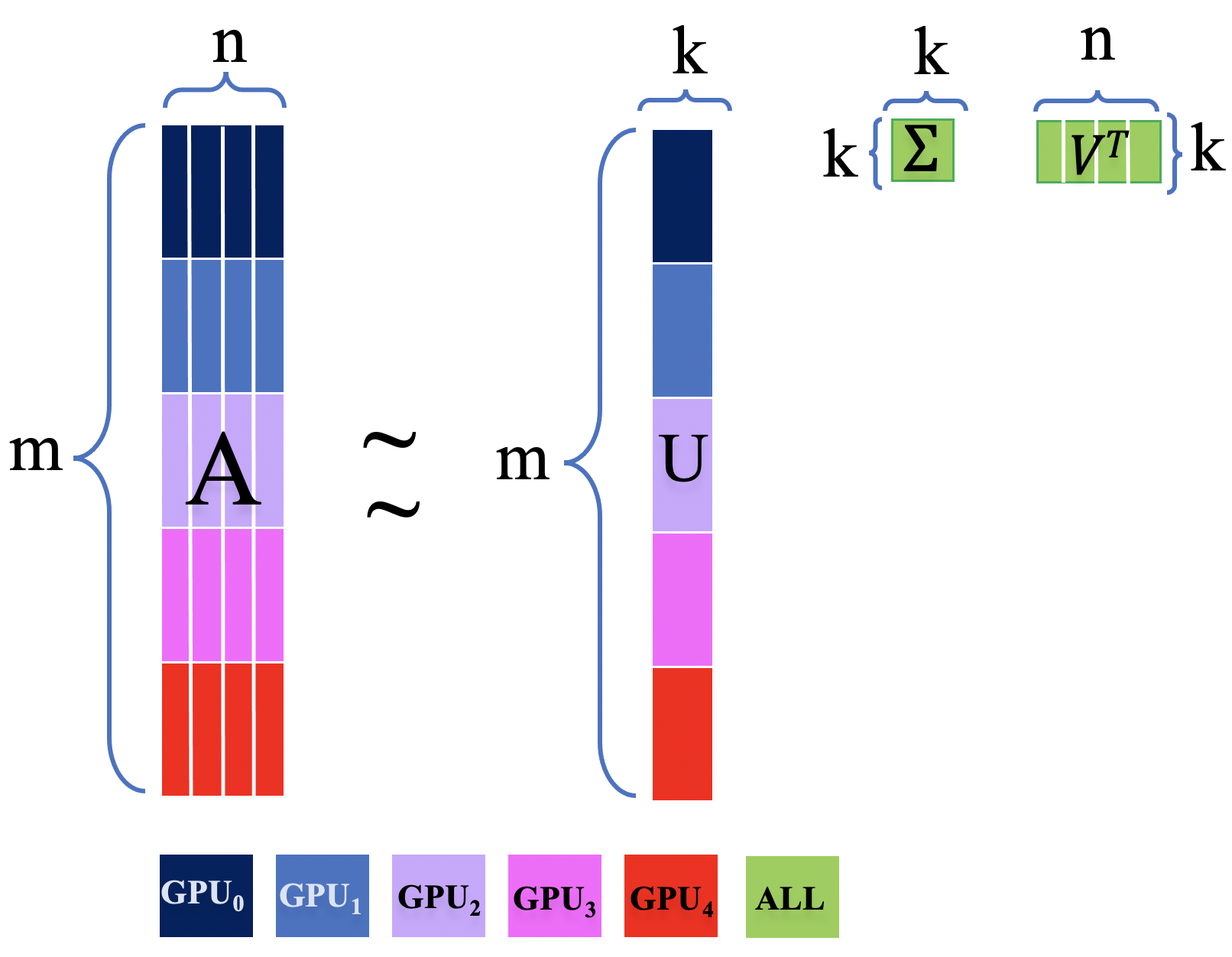}
        \subcaption{Problem partition and data locality   \label{fig:data_locality}}
    \end{minipage}
        \hspace{2em}
    \begin{minipage}[b]{0.28\linewidth}
        \includegraphics[width=.8\linewidth]{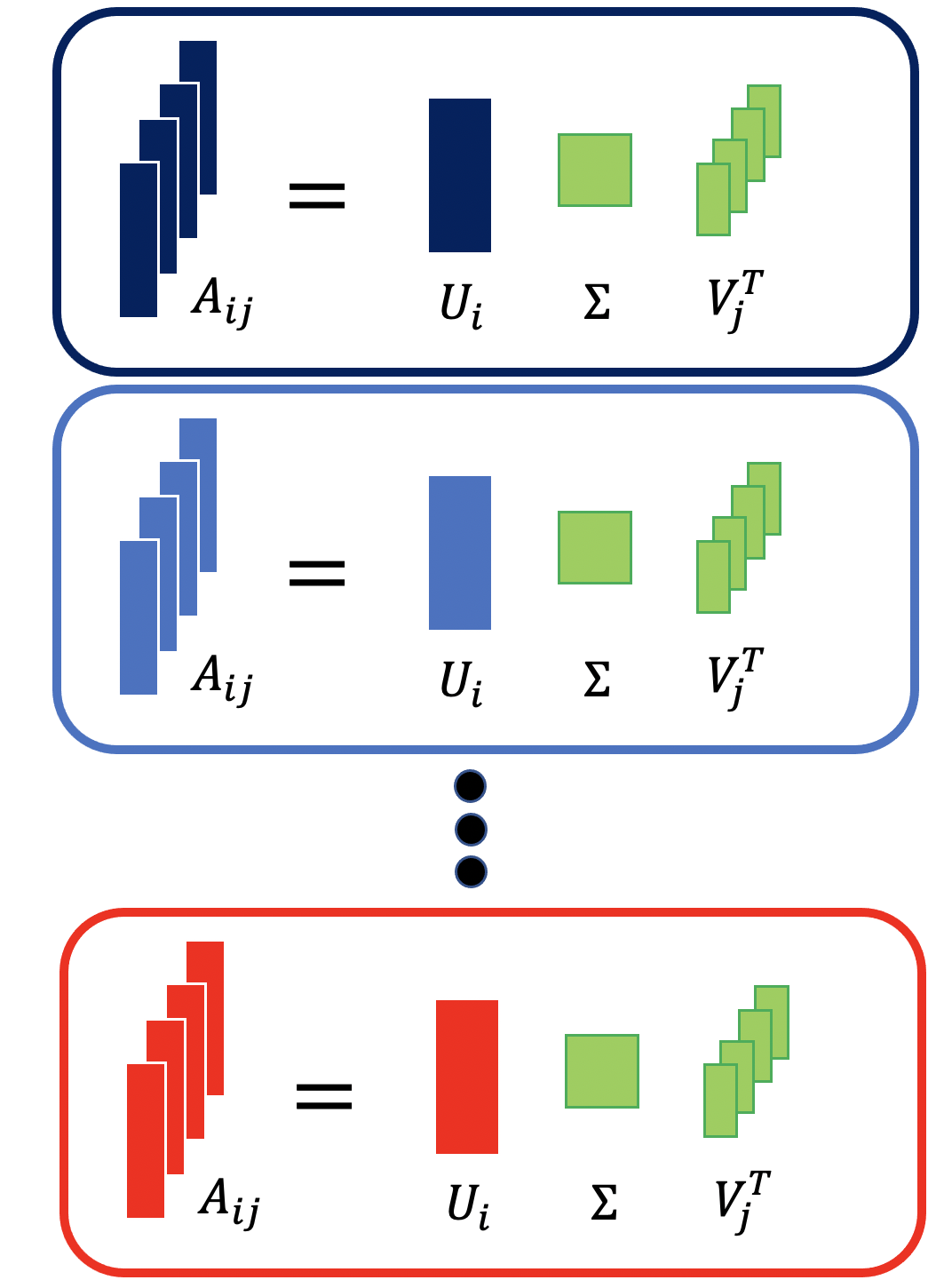}
        \subcaption{Parallel batch/tile implementation \label{fig:parallel_implementattion}}
    \end{minipage}
    \quad
    \caption{Illustration of the truncated SVD of $\mat{A}$ in a distributed setting consisting of $N=5$ GPUs. Row partition of problem is illustrated in (a) and data locality is indicated by colored zones, coded with the given legend at the bottom. Green color (all) indicates a replication of the same data across all GPUs. Vertical solid lines illustrate the segmentation of local $\mat{A}$ into 4 batches/tiles, and parallel implementation is illustrated in (b).} \label{fig:diributed_SVD}
    \vspace{-1.3em}
\end{figure*}
\section{Related work}


SVD is an integral component of many machine learning algorithms involving dimensionality reduction, data compression\cite{bhattarai2020distributed}, knowledge graphs\cite{bhattarai2022distributed}, and information retrieval\cite{pyDNMFk}. With the growing volume of data from social networks, autonomous driving, and simulations, regular SVD algorithms cannot perform the decomposition over a single processor due to memory constraints. To address this memory bottleneck, various out-of-memory algorithms and distributed algorithms have been devised in either CPU or GPU hardware.\cite{dongarra2018singular} provides an overview of different large-scale SVD frameworks designed to decompose large-scale datasets. The work in papers \cite{3,kabir2017framework,dongarra2017framework,rabani2001out} demonstrate the out-of-memory SVD implementation for CPU hardware whereas papers \cite{haidar2017out,boukaram2018batched,lu2020reducing,gates2018accelerating} demonstrate the implementation for the GPU hardware. A majority of the existing \emph{out-of-memory} (OOM) and distributed implementation of SVD emphasize computing only the singular values, which is insufficient for applications such as tensor networks, which need accurate computation of a substantial number of singular vectors. In addition, a majority of them can only decompose a matrix of size $m\times n$ when $m\gg n$ or $m\ll n$ as the data partitioning is done along one axis. However, most real-world datasets such as knowledge graphs tend to have symmetrical matrix shapes, i.e., $m\approx n$, and one can't apply such distributed/OOM implementations.
Furthermore, most of the OOM and the distributed implementations exist in the CPU. Still, the data size becomes a bottleneck given the hardware constraint for these implementations. Also, the distributed implementations  for GPU reported in \cite{5,zampini2022h2opus} do not report the scalability for very large datasets. Hence it is vital to take advantage of both the out-of-memory ability combined with the distributed implementation to decompose matrices of unprecedented sizes at an accelerated pace on GPU. This approach has recently been utilized in our distributed out-of-memory non-negative matrix factorization (NMF) paper\cite{boureima2022distributed} for complexity 0 out-of-memory situations(discussed later). In such distributed out-of-memory implementation, the distributed axis splits one of the co-factor matrices.
In contrast, the batched axis splits the other co-factor matrix enabling the decomposition of massive sparse datasets. Furthermore, our distributed implementation is optimized with NCCL \cite{awan2016efficient} communicator for accelerated inter and intra GPU communication. In contrast, CUDA streams accelerate the data movement between GPU and CPU for out-of-memory implementation.   
\section{Rationale for an algorithm for the Out-of-memory distributed SVD}
The proposed implementation of SVD is for heterogeneous HPC systems, with the ability to handle OOM scenarios, where the data is too big to be cached on combined GPU memory. A serial algorithm for the truncated SVD is given in Algorithm \ref{alg:svd_trunc}, from which we can estimate the memory complexity of SVD for a given matrix $\mat{A}$ of size $S_{\mat{A}}$ (in bytes) to be $S_{SVD} \approx 4 \times S_{\mat{A}}$, assuming $k << (m,n)$. Starting counting at line 6 of Algorithm \ref{alg:svd_trunc}, two buffers of size $S_A$ are required for the copy $\mat{X}=\mat{A}$ ( $\sim 2 \times S_{\mat{A}}$ total), the memory utilization will peak at line 8 of Algorithm \ref{alg:svd_trunc}, where two additional buffers of size $S_A$ will be required to compute the residual $\mat{X}$ = $\mat{X} - \mat{U}[:l]diag(\mat{\sigma}[:l])\mat{V}[:l]^T$ (totalling $\sim 4 \times S_{\mat{A}})$. Note that when $\mat{A}$ is sparse, choosing to represent $\mat{A}$ in a sparse format, such as Compressed Sparse Row(CSR) or Coordinate Format(COO), can dramatically reduce the memory required to store $\mat{A}$. This memory reduction can skew the calculation of $S_{SVD}$ because the resulting co-factors $\mat{U}$ and $\mat{V}$ are dense, and most importantly, the residual $\mat{X}$ is now dense with the same shape as $\mat{A}$, and now $S_{SVD} \approx 2 \times S_A$, where $S_A$ is the size of the dense representation of $\mat{A}$. 

When performing SVD on GPUs, OOM situations can arise in various scenarios with different degrees of complexity. We distinguish three main degrees of complexity for OOM scenarios. Scenarios of complexity \emph{degree 0} concern practical problems where the input data $\mat{A}$ and its co-factors $\mat{U}$, $\mat{\sum}$, and $\mat{V}$ can easily be stored on GPU memory, but an explosion of memory requirement occurs when computing intermediate results. This is often the case when computing the intermediate results $\mat{X} = \mat{X}  - \mat{U}\mat{\Sigma}\mat{V}^T$ (line 8 of Algorithm \ref{alg:svd_trunc}), when $\mat{A}$ is a large sparse matrix of very low density and the factors are dense. The matrix $\mat{X}$ resulting from the operation becomes dense and very likely impossible to store on GPU. For instance, if $\mat{A}\in \mathbb{R}^{10^6 \times 10^6}$ is a sparse matrix, with density of $\delta \approx 10^{-3}$, the size of $\mat{A}$ in dense format, in single precision, is $S_{\mat{A}}\approx 4 TB$, however representing $\mat{A}$ in CSR sparse format can lower the size of $\mat{A}$ down to $S_s \sim S_A \times \delta \approx 4GB$, consequently $S_{SVD} \approx 2 \times S_{\mat{A}}\approx 4TB$. Assuming very small k, $\mat{A}$ and all co-factors can be stored on GPU, however the calculation of the residual $\mat{X}$ from  $\mat{X} = \mat{X}  - \mat{U}\mat{\Sigma}\mat{V}^T$ would still require a whopping $\sim 8000 GB$ of GPU memory (line 9 of Algorithm \ref{alg:svd_trunc}) make this scenario a degree 0 OOM complex problem. Below we propose a solution to address this bottleneck, where we avoid the direct matrix-matrix multiplications that would result in such a large dense matrix. Instead, we perform a series of matrix-vector computations to minimize the overall computation and memory cost. 
 
 A higher degree of complexity, \emph{degree 1}, arises in cases where matrix $\mat{A}$ and at most two of its co-factors cannot be cached on GPU memory; this is typically the case when dealing with a large $\mat{A}$ that is dense or sparse with high density. Scenarios of complexity \emph{degree 2} are the most complex and consist of practical cases where neither $\mat{A}$, nor its co-factors can be stored on GPU memory. Note that more complexity can arise in cases where data cannot fit on host RAM memory, but that still is of degree 2 as the complexity here is measured with respect to GPU RAM memory.
 In other words, in \emph{degree 0} complex scenarios, all the data can be cached on GPU; in \emph{degree 1} complex scenarios, the data can partially be cached on GPU, and in \emph{degree 2} complex scenarios, none of the data can be cached on GPU.
 
 The treatment of \emph{degree 2} complex scenarios is out of the scope of this study. We propose \emph{tiling} and \emph{batching} techniques to deal with OOM problems for degree 0 and degree 1. 
 Both \emph{Tiling} and \emph{batching} are block-based computation techniques that operate the same way; however, in the interest of clarity, we differentiate the two base on the employed data transfer pipelines. \emph{Tiling} takes place on GPU and employs GPU data transfer pipelines such as global memory to shared memory data transfer pipelines, while \emph{batching} takes place between host (CPU) and device (GPU) and uses the host to device data transfer pipelines and vice-versa. Optimization of \emph{Tiling} techniques relies on GPU architecture and features such as memory speed, available VRAM, etc. In contrast, the optimization of \emph{batching} techniques depends more on the speed of busses connecting host and device, such as PCIe or NV-Link bus speeds, as well as the number of CUDA streams, which can be exploited in an asynchronous approach to overlap data transfer and compute operation to hide data copy latency. We adopt \emph{tiling} techniques to deal with problems of complexity \emph{degree 0}, and \emph{batching} technique to handle \emph{degree 1} complex problems. In extreme cases, we will complement \emph{batching} by \emph{tiling} to further reduce memory footprint.
 Bellow, we discuss our implementation and design choices.

\section{Out of Memory algorithm design for large sparse datasets}
\label{sec:OOM}
The power method approach to performing truncated SVD is a popular approach that is based on repeatedly performing a linear projection on an initial vector until it converges to the direction of each desired eigenvector. A serial version of the truncated SVD algorithm is given in Algorithm~\ref{alg:svd_trunc}, which relies on Algorithm~\ref{alg:svd_1d} for estimating $k$ singular vectors where each singular vector is estimated one at a time. The algorithm's major computational/memory bottlenecks are in the computation of the Gram matrix (i.e., lines 7 and 9) in  Algorithm~\ref{alg:svd_1d}. In addition, the computation of line 9 in Algorithm~\ref{alg:svd_trunc} requires the computation of the product of the SVD factors requiring additional memory. The distributed realization of the Gram matrix computation is presented in Algorithm~\ref{alg:dist_gram}. We utilize this distributed gram-based realization of Algorithm~\ref{alg:svd_1d} for dense datasets. However, due to the bottleneck associated with computing the gram of residual matrix from very large dense cofactors for the sparse dataset, we avoid this distributed realization for the sparse dataset and propose a new realization discussed next. We reduce memory complexity using the analytical derivation below:\\
From the line 8 of Algorithm~\ref{alg:svd_trunc}, we have the following expression.
\begin{equation*}
  \mat{X}'= \mat{X} - \mat{U}[:l] diag(\mat{\sigma}[:l])\mat{V}[:l]^T   
\end{equation*}

let $\mat{U} = \mat{U}[:l]$, $\mat{V} = \mat{V}[:l]$ and $\mat{\Sigma} = diag(\mat{\sigma}[:l])$. 

Then for $m>n$, we have
\begin{equation}
\label{eqn:expansion1}
\begin{split}
    \mat{B} &= \mat{X}'^T \mat{X}'  = (\mat{X} - \mat{U}\mat{\Sigma}\mat{V}^T)^T (\mat{X} - \mat{U}\mat{\Sigma}\mat{V}^T) \\
            & = (\mat{X}^T - \mat{V} \mat{\Sigma}^T\mat{U}^T)
            (\mat{X} - \mat{U}\mat{\Sigma}\mat{V}^T) \\
            &= \mat{X}^T \mat{X} -\mat{V} \mat{\Sigma}^T\mat{U}^T \mat{X} -\mat{X}^T \mat{U}\mat{\Sigma}\mat{V}^T +\mat{V} \mat{\Sigma}^T\mat{U}^T \mat{U}\mat{\Sigma}\mat{V}^T  \\
            &= \mat{X}^T \mat{X} -\mat{V} \mat{\Sigma}^T\mat{U}^T \mat{X} -\mat{X}^T \mat{U}\mat{\Sigma}\mat{V}^T +\mat{V}\mat{\Sigma}^2 \mat{V}^T,
\end{split}
\end{equation}
since $\mat{U}^T \mat{U} = I$ and $\mat{\Sigma} \mat{\Sigma}^T = \mat{\Sigma}^2$ as $\mat{\Sigma}$ is a diagonal matrix. 

Still, Equation~\ref{eqn:expansion1} is computationally and memory expensive due to matrix-matrix multiplications and requirement to store $n\times n$ dense matrix. So we avoid the direct computation of Equation~\ref{eqn:expansion1}, but instead, we multiply the expression with vector $\mat{v}_0$ as shown in line 11  of Algorithm~\ref{alg:svd_1d}. So we can represent 
\begin{equation}
\label{eqn:expansion2}
\begin{split}
\mat{v}_1 &= \mat{B} \mat{v}_0 \\
          &= (\mat{X}^T \mat{X} -\mat{V} \mat{\Sigma}^T\mat{U}^T \mat{X} -\mat{X}^T \mat{U}\mat{\Sigma}\mat{V}^T +\mat{V}\mat{\Sigma}^2 \mat{V}^T) \mat{v}_0 \\
          &= \mat{X}^T \mat{X} \mat{v}_0 - \mat{V} \mat{\Sigma}^T\mat{U}^T \mat{X} \mat{v}_0 -\mat{X}^T \mat{U}\mat{\Sigma}\mat{V}^T\mat{v}_0 +\\ 
          &     \mat{V}\mat{\Sigma}^2 \mat{V}^T \mat{v}_0, 
 \end{split}
\end{equation}
where each product is computed in the right-to-left order.
Similarly, for $m<n$, we have:
\begin{equation}
\label{eqn:expansion3}
\begin{split}
\mat{v}_1 &= \mat{X}\mat{X}^T\mat{v}_0 -\mat{U} \mat{\Sigma}^T\mat{V}^T \mat{X}^T\mat{v}_0-\mat{X} \mat{V}\mat{\Sigma}^T\mat{U}^T\mat{v}_0 +\\ 
          &     \mat{U}\mat{\Sigma}^2 \mat{U}^T \mat{v}_0.
\end{split}
\end{equation}
\begin{algorithm}
\scriptsize
    \caption{\small dist\_gram($\mat{X},n_b,N$)}
    \label{alg:dist_gram}
    \textbf{Require: $\mat{X}$} $\in \mathbb{R}_{+}^{m \times n}$.
    \textbf{Require: $\mat{X}$} distributed across $N$ GPUs where  $\mat[][ij]{X} \in \mathbb{R}^{m/N \times n}$ if $m>n$ otherwise  $\mat[][ij]{X} \in \mathbb{R}^{m \times n/N}$. Locally to each GPU, $\mat[][ij]{X}$ can be split into $n_b$ batches following a \emph{co-linear} batching strategy where the batch size is $b_s = max(m,n)/(N*n_b)$, or an \emph{orthogonal} batching strategy where the batch is $b_s = min(m,n)/n_b$. 
	\begin{algorithmic}[1]
	\State $m_i,n_j$ = shape($\mat[][ij]{X}$) \Comment{Depending upon $m>n$ or $m<n$, the GPU partitions $m$ or $n$ yielding index $i$ or $j$ at the first stage and than  collinear or orthogonal batching strategy dictates index at second stage}
    
    \State Initialize local array $\mat{B}$, SQUEUE, a queue of CUDA-streams of size $qs$, and tile size $q = min(m,n)/N$ 
     \For {$j$ in $n_b$}
     \State $j_o,j_1$ = $j*b_s,(j+1)*b_s$
     \State $\mat{A}$ = $\operatorname{H2D}(\mat{X}[:, j0:j1])$ \Comment{H2D stands for copy from host to GPU}
     \For {$i$ in $j+1$}
     \State SQUEUE $->$ stream \Comment{De-queue stream from SQUEUE}
     \State $\mat{AT}$= $A^T$
     \State $i_0, i_1$ = $i*b_s, (i+1)*b_s$
    \State  $\mat{A} = \operatorname{H2D}( \mat{X}[:, i_0:i_1] )$ 
    \State $\mat{B} = \mat{AT}@\mat{A}$ \Comment{Compute Gram locally}
    \State $\mat{B}$ = Reduce\_sum($\mat{B}$, root) \Comment{Reduce B along the communicator with root=$j$ for lower diagonal elements and with root=$i$ otherwise}
 \State  stream $->$ SQUEUE  \Comment{En-queue stream back into SQUEUE}
 \EndFor
 \EndFor
 \State return $\mat{B}$
\end{algorithmic} 

\end{algorithm}

Equations~\ref{eqn:expansion2} and \ref{eqn:expansion3} mean that we significantly reduce memory complexity by avoiding the direct calculation of the residual $\mat{X}$ as well as the calculation of the Gram $\mat{B}=\mat{X}'^T\mat{X}$, and instead computing directly the  $k^{th}$ singular vector skipping lines 6-9 in Algorithm ~\ref{alg:svd_1d}. Evaluating the expression in Equations~\ref{eqn:expansion2} or \ref{eqn:expansion3}  right to the left would replace the matrix-matrix multiplications with a series of matrix-vector multiplications, thus significantly lowering memory complexities. The newly devised algorithm's distributed implementation is presented in Algorithm~\ref{alg:dist_compute_v}. This can, however, increase the time complexity of the degree 1 OOM complex problem due to the need to batch in each iteration any of the matrices that are not cached on GPU. While this data movement cost is reasonable for a sparse dataset, the cost can be a significant bottleneck for dense datasets due to the considerable data movement for OOM implementation. Consequently, a design trade-off between time and memory complexity needs to be assessed based on the available hardware, i.e., this can be a viable option with modern GPUs with high-bandwidth SXM/NVLink interfaces. 
\begin{algorithm}
\scriptsize
    \caption{\small dist\_Compute\_v0($\mat{X},\mat{U},\mat{\Sigma},\mat{V},n_b,N$)}
    \label{alg:dist_compute_v}
    \textbf{Require: $\mat{X}$} $\in \mathbb{R}_{+}^{m \times n}$, $\mat{U} \in \mathbb{R}_{+}^{m \times k}$,$\mat{V} \in \mathbb{R}_{+}^{m \times k}$ and $\mat{\Sigma} \in \mathbb{R}_{+}^{k \times k}$\\
    \textbf{Require: $\mat{X}$} distributed across $N$ GPUs where  $\mat[][ij]{X} \in \mathbb{R}_{+}^{\frac{m}{N} \times \frac{n}{n_b}}$ if $m>n$ otherwise  $\mat[][ij]{X} \in \mathbb{R}^{\frac{m}{n_b} \times \frac{n}{N}}$. 
    If $m>n$, then $\mat{U}$ is distributed across N GPUs where $\mat{U}_i \in \mathbb{R}_{+}^{\frac{m}{N}\times k}$ while $\mat{V}$ and $\mat{\Sigma}$ are the same across every GPU where $\mat{V}$ is batched  such that $\mat{V}_j \in \mathbb{R}_{+}^{ \frac{n}{n_b}\times k}$ . Similarly, if $m<n$, then   $\mat{V}$ is distributed across N GPUs where $\mat{V}_i \in \mathbb{R}_{+}^{\frac{n}{N}\times k}$ while $\mat{U}$ and $\mat{\Sigma}$ are the same across every GPU where $\mat{U}$ is batched  such that $\mat{U}_i \in \mathbb{R}_{+}^{\frac{m}{n_b}\times k}$
	\begin{algorithmic}[1]

    \If {$m>=n$}
    	\State  $I=m/N$ and $b=n/n_b$
    \State $(\mat{Xv_0})_{ij}$ = $\mat{X_{ij}} @ (\mat{v_0})_j$ \Comment{Multiply a batch of $\mat{X}$ of size $I\times b$ with a batch of $\mat{v_0}$ of size $b \times 1$ so that a batch of $\mat{Xv_0}$ is of size $I$  for $n_b$ batches. }
    \State  $(\mat{Xv_0})_{i}$ = $\sum_{j=1}^b(\mat{Xv_0})_{ij}$ \Comment{Reduction along batches so that $(\mat{Xv_0})_{i}$ is of size $I\time 1$}
    \State $(\mat{X^TXv_0})_{i,j}$ = $\mat{X}_{ij}^T @ (\mat{Xv_0})_{i}$ \Comment{($\mat{X^TXv_0})_{i,j}$ is of size $b\times 1$ for $n_b$ batches for N GPUS.}
     \State $(\mat{X^TXv_0})_{j}$ = $\sum_{i=1}^N (\mat{X^TXv_0})_{i,j}$ \Comment{All reduce sum along the GPUs i.e. $i$ axis such that $(\mat{X^TXv_0})_{j}$ is of size $b\times 1$ for $n_b$ batches along each GPU.} \label{expr1}
   \State $(\mat{U^TXv_0})_{i}$ = $\mat{U}_{i}^T @ (\mat{Xv_0})_{i}$ \Comment{($\mat{U^TXv_0})_{i}$ is of size $k\times 1$} 
   \State $\mat{U^TXv_0}$ = $\sum_{i=1}^N (\mat{U^TXv_0})_{i}$ \Comment{All reduce sum along the GPUs so that $\mat{U^TXv_0}$ is of size $k\times 1$ and same for all GPUs}
      \State $\mat{\Sigma^T U^TXv_0}$ = $\mat{\Sigma}^T @ \mat{U^TXv_0}$ \Comment{$\mat{\Sigma^T U^TXv_0}$ is of size $k\times 1$}
            \State $(\mat{V\Sigma^T U^TXv_0})_{j}$ = $\mat{V}_j @ \mat{\Sigma^T U^TXv_0}$ \Comment{$\mat{V\Sigma^T U^TXv_0}$ is of size $b\times 1$ for $n_b$ batches.} \label{expr2}
    \State $(\mat{V^Tv_0})_{j}$ = $\mat{V_{j}} @ (\mat{v_0})_j$ \Comment{Multiply a batch of $\mat{V}^T$ of size $k\times b$ with a batch of $\mat{v_0}$ of size $b \times 1$ so that a batch of $\mat{V^Tv_0}$ is of size $k$  for $n_b$ batches. }
     \State  $\mat{V^Tv_0}$ = $\sum_{j=1}^b(\mat{V^Tv_0})_{j}$ \Comment{Reduction along batches so that $\mat{V^Tv_0}$ is of size $b\time 1$, which is the same for every GPU.}
    \State $\mat{\Sigma V^Tv_0}$ = $\mat{\Sigma} @ \mat{V^Tv_0}$ \Comment{$\mat{\Sigma V^Tv_0}$ is of size $k \times 1$}
   \State $(\mat{U\Sigma V^T v_0})_i$ = $\mat{U}_i @ \mat{\Sigma V^Tv_0}$ \Comment{$(\mat{U\Sigma V^T v_0})_i$ is of size $I \times 1$}
 \State $(\mat{X^T U\Sigma V^T v_0})_{ij}$ = $\mat{X}^T_{ij} @ (\mat{U\Sigma V^T v_0})_i$ \Comment{$(\mat{X^T U\Sigma V^T v_0})_{ij}$ is of size $b \times 1$ for $n_b$ batches different values on each GPU.}  
  \State $(\mat{X^T U\Sigma V^T v_0})_{j}$ = $\sum_{i=1}^N (\mat{X^T U\Sigma V^T v_0})_{ij}$ \Comment{$(\mat{X^T U\Sigma V^T v_0})_{j}$ is of size $b \times 1$ for $n_b$ batches same values on each GPU.} \label{expr3}
 \State $\mat{\Sigma^T V^Tv_0}$ = $\mat{\Sigma} @ \mat{\Sigma V^Tv_0}$ \Comment{$\mat{\Sigma^T V^Tv_0}$ is of size $k \times 1$}
    \State $(\mat{X^T U\Sigma V^T v_0})_{j}$ = $\mat{V}_j @ \mat{\Sigma^T V^Tv_0}$ \Comment{$(\mat{V\Sigma^T V^Tv_0})_j$ is of size $b \times 1$  for $n_b$ batches.} \label{expr4}
\State $(\mat{v_1})_j$ = $(\mat{X^TXv_0})_{j}$ - $(\mat{V\Sigma^T U^TXv_0})_{j}$ - $(\mat{X^T U\Sigma V^T v_0})_{j}$ + \textbf{$(\mat{X^T U\Sigma V^T v_0})_{j}$} \Comment{Using \ref{expr1},\ref{expr2},\ref{expr3} and \ref{expr4}, compute $(\mat{v_1})_j$ for $n_b$ batches}
    \EndIf
    \State return $(\mat{v_1})_j$
         	\end{algorithmic} 
\end{algorithm}
\section{pyDSVD for distributed heterogeneous systems}
\subsection{Implementation of SVD for distributed heterogeneous systems}
\label{sec:Implementation}

The proposed implementation of SVD for distributed heterogeneous systems partitions large SVD problems into smaller distributed problems using strong considerations for data locality, such as avoiding inefficiencies associated with communication (data transfer) in the distributed system. Additional trade-offs are considered in OOM scenarios; for instance, to reduce communication, it is sometimes better to replicate data over the distributed grid, while other times, it is acceptable to use batching/tiling techniques that can increase communication latency to lower the memory footprint. Bellow, we discuss the problem partition strategies in subsection(\ref{subsec:dist_partition}), followed by a discussion of our tiling and batching approaches in subsection(\ref{subsec:dist_batching}).

\begin{figure}[t!]
\centering
    \begin{minipage}[b]{0.47\linewidth}
        \includegraphics[width=.8\linewidth]{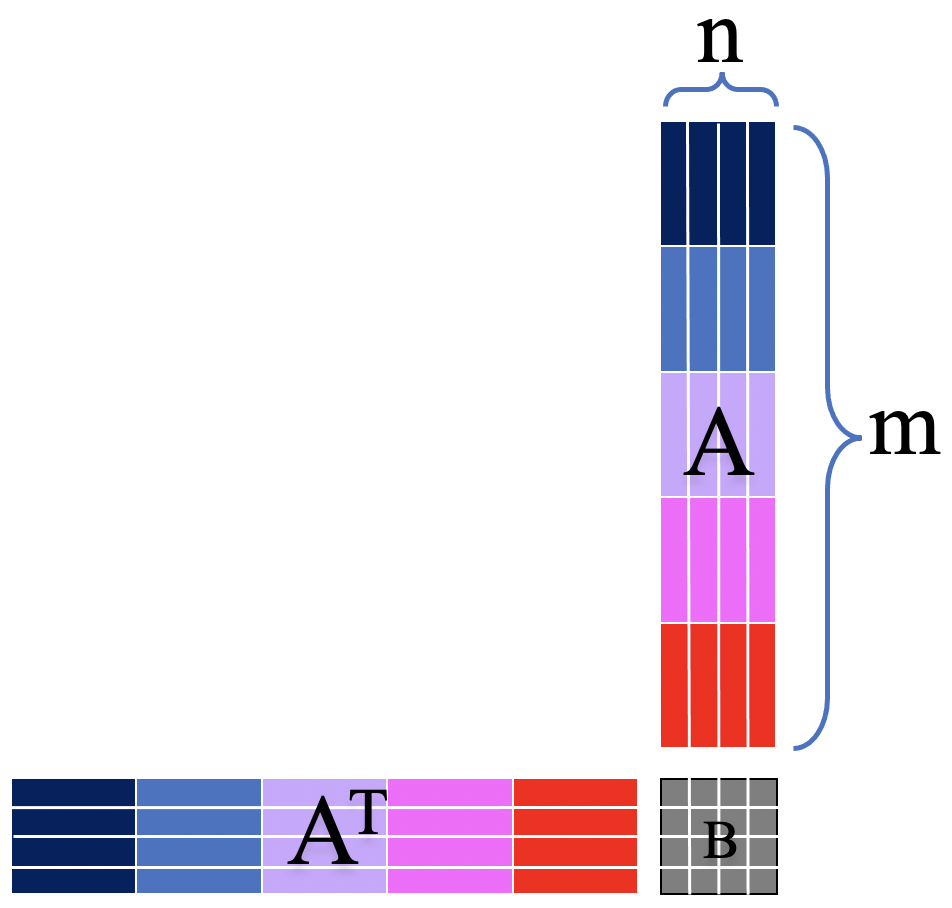}
        \subcaption{Gram $\mat{B}=\mat{A^T}\mat{A}$ \label{fig:ATA}}
    \end{minipage}
    \begin{minipage}[b]{0.47\linewidth}
        \includegraphics[width=.8\linewidth]{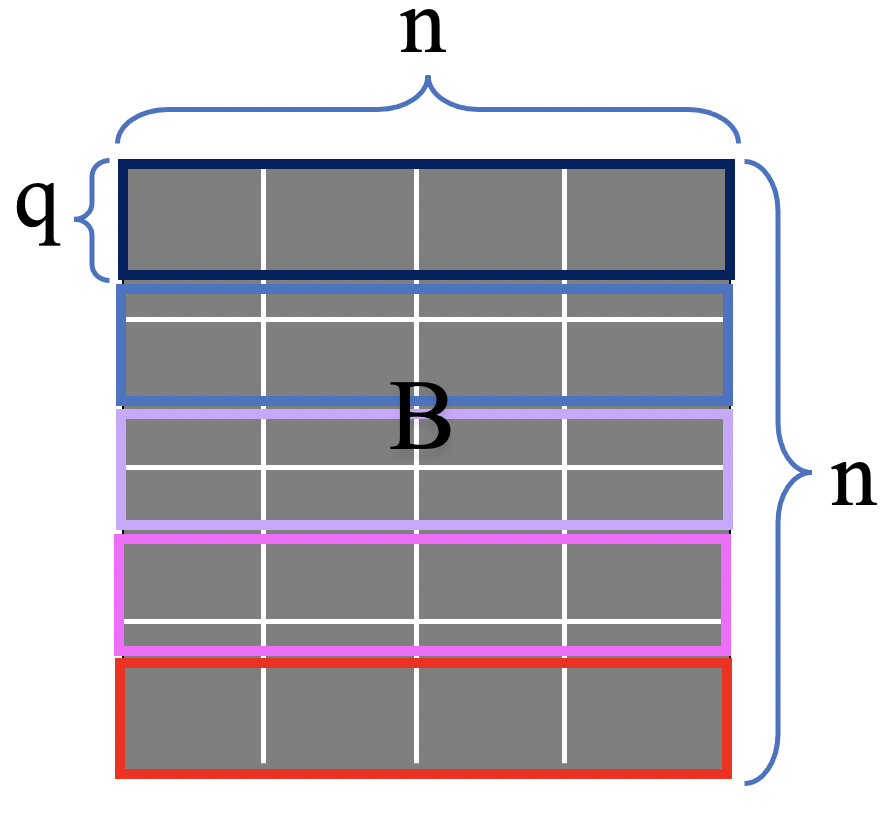}
        \subcaption{Distributed $\mat{B}$ \label{fig:B}}
    \end{minipage}
 
    \begin{minipage}[b]{0.47\linewidth}
        \includegraphics[width=.8\linewidth]{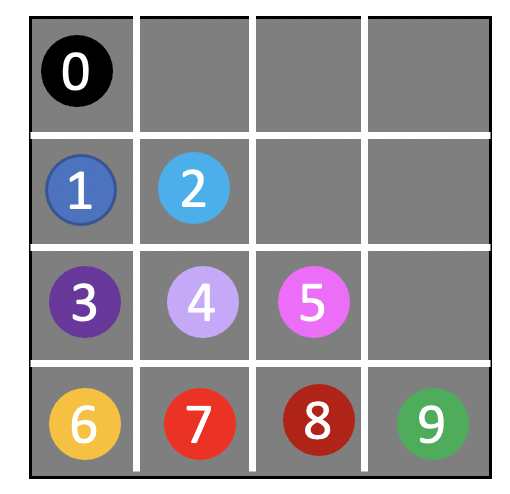}
        \subcaption{$\mat{B}$ computation task list \label{fig:task_lisk}}
    \end{minipage}
    \begin{minipage}[b]{0.4\linewidth}
        \includegraphics[width=.8\linewidth]{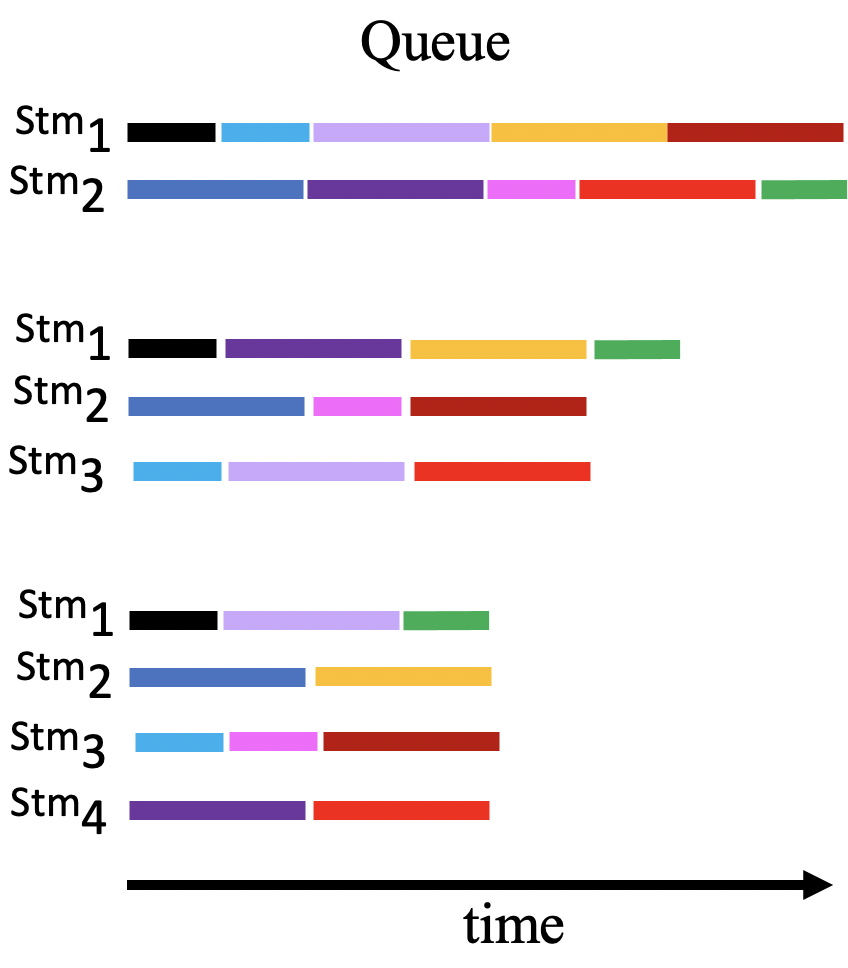}
        \subcaption{$\mat{B}$ computation task scheduling \label{fig:task_scheudule}}
    \end{minipage}

    \caption{ Illustration of distributed Gram product  $\mat{B}=\mat{A^T}\mat{A}$ in (a) for $\mat{A}$  distributed across 5 GPUs, color-coded with same colors shown in the legend of Figure~\ref{fig:data_locality}. Vertical solid lines illustrate the segmentation of local $\mat{A}$ into 4 batches and a possible distributed partition of $\mat{B}$ (when ${\mat{B}}$ is too big) across different GPUs in (b). Chosen task organization for computing $\mat{B}$ is shown in (c) and the corresponding asynchronous task scheduling for different queue sizes is shown in (d).} \label{fig:B_batching}
    \vspace{-1.3em}
\end{figure}

\subsection{Partition of computational work on the distributed HPC system}
\label{subsec:dist_partition}
Our implementation considers two one-dimensional data partition strategies based on the shape of matrix $\mat{A}$ ($m\times n$). A column (vertical) partition, \emph{CSVD} employed when $n > m$, and a row (horizontal) partition, \emph{HSVD}, used otherwise.

Assuming a distributed system with $N$ GPUs, where each GPU is indexed by its global rank $g_{ID}$. In the \emph{RSVD} approach illustrated in Figure~\ref{fig:data_locality}, the $i^{th}$ GPU with $g_{ID}=i$ will work on array partitions $\mat{A}[i_0:i_1, :]$, $\mat{U}[i_0:i_1, :]$, $\mat{\Sigma}$, and $\mat{V}$, where $i_0=i \times I$, $i_1=(i+1) \times I$, and $I=m/N$(partition size). Each GPU gets a full copy of $\mat{\Sigma}$  and $\mat{V}$ ($\mat{\Sigma}$ and $\mat{V}$ are replicated) and a unique partition of $\mat{A}$ and $\mat{U}$. This translates into a segmentation of arrays $\mat{A}$ and $\mat{U}$ on global memory, as we illustrate for the case $N=5$ in Figure~\ref{fig:data_locality}. The colored zones indicate local partitions of $\mat{A}$ on each GPU in global memory (data locality), and consequently help conceptualize  communication  requirements whenever information is exchanged from one zone to another.

\subsection{Out-of-memory implementation}
\label{subsec:dist_batching}

In out-of-memory situations where the available GPU memory $S_{G}$ is insufficient ($S_{G} < S_{SVD}$), light arrays are cached on GPU memory, and heavier arrays are kept on host memory.  We further split the local blocks of data into smaller batches/tiles in a way that is either collinear to the direction of the larger dimension $\mat{A}$, i.e., $m$ in RSVD  where $m >n$, or in a way that is orthogonal to the direction of $m$. In the former method, we talk about adopting a \emph{collinear} batching/tiling technique; in the latter, we talk about adopting an \emph{orthogonal} batching/tiling technique. 
To illustrate, let $b_s$ be a batch size control parameter, and let's assume we have a $RSVD$ ($CSVD$) partition scenario. The number batches in an orthogonal batching technique is then given by $n_B = n/b_s$ ($n_B = m/b_s$), and $\mat{U}[I,:]$ $(\mat{V}[:,J])$ is cached on GPU memory whereas, heavier arrays $\mat{A}[I, b_0:b_{1}]$ ($A[b_0:b_{1}, J]$) and $\mat{V}[:, b_0:b_{1}]$ ($\mat{U}[b_0:b_{1}, :]$) are batched to their respective GPUs as needed, such that for the $b^{th}$ batch, $b_0=b \times b_s$ and  $b_1=(b+1) \times b_s$. 
Cartoons in Figure~\ref{fig:parallel_implementattion} illustrate the orthogonal tiling strategy of a $RSVD$ partition of $\mat{A}$ among 5 GPUS. Data locality is indicated by colored zones, coded with the legend at the bottom. $\mat{V}$ and $\mat{\Sigma}$ are replicated on all GPUs, as indicated by \emph{green} color. Vertical solid lines illustrate the segmentation of local data into batches/tiles, as is the case for $\mat{A}$ and $\mat{V}$. Note that these scenarios assume a problem of OOM complexity of \emph{degree 0} because all data blocks are on GPU memory. In \emph{degree 1} $\mat{A}$ and $\mat{V}$ would have been on host RAM memory.

Figure~\ref{fig:B_batching} and Algorithm~\ref{alg:dist_gram} highlights the important aspects of computing the Gram product $\mat{B}=\mat{A}^T\mat{A}$, in a distributed setting, and possibly for OOM scenarios. In this illustration, the distributed HPC system has 5 GPUs among which $\mat{A}$ is distributed ($\mat{A}$ may or may not be cached on GPU), data locality is color coded same as in subsection (\ref{subsec:dist_partition}), and vertical white lines delineates batch/tile boundaries. The distributed and batched/tilled Gram product $\mat{B}=\mat{A}^T\mat{A}$ is illustrated in 
Figure~\ref{fig:ATA}, and the resulting distributed matrix $\mat{B}$ is illustrated in Figure~\ref{fig:B}. The distributed matrix $\mat{B}$ is here of size $n \times n$. It may not be possible to replicate on the different GPUs in OOM scenarios, in which case $\mat{B}$ can be distributed as indicated by the colored rectangles in Figure~\ref{fig:B}, showing a possible distribution of $\mat{B}$ among the five different GPUs. Note that the white and colored lines do not overlap because the number of batches $nb$ is not divisible by the number of GPU $N$. $\mat{B}$ can be computed using $n_T= nb \times nb = 4 \times 4 = 16$ independent tasks, each with an independent results $B_{ij}$ of shape $b_s \times b_s$ represented by the grey squares forming $\mat{B}$. In \emph{degree 1} complex problem, each task $T_{ij}$ involves a \emph{H2D($\mat{A}_i$, $\mat{A}_j$)} of the local batches $\mat{A_i}$ and $\mat{A}_j$ followed by the computation of $\mat{A_i}^T$, then $\mat{B}_{ij}=\mat{A}^T_i \times \mat{A}_j$ and subsequently, the copy \emph{D2H($\mat{B}_{ij}$)} back to the host if $\mat{B}$ is not cached on device. Each Task can be queued to a non-default CUDA stream $Stm_{ij}$ as the tasks are independent, for an asynchronous calculation of $\mat{B}$. Controlling the queue size $qs$ allows controlling the number of concurrent tasks running on each GPU, which in turn allows controlling the GPU memory utilization since we can also control the batch size $b_s$. Further, note that $\mat{B}$ can be computed with $nr_T = 10 < n_T$ fewer tasks if we were to reuse data already uploaded to the GPU: The lower triangular part of $\mat{B}$ is symmetric to the upper triangular part by transposition. Consequently each task $T_{ij}$ computing non-diagonal $\mat{B}_{ij}$ can save an extra \emph{H2D($\mat{A}_j$, $\mat{A}_i$)},  and  \emph{D2H($\mat{B}_{ji}$)} by computing the symmetrical $\mat{B}_{ji}=[\mat{A}^T_i \times \mat{A}_j]^T=\mat{A}^T_j \times \mat{A}_i$, which is then sent to the GPUs responsible for storing it. Figure~\ref{fig:task_lisk} illustrates the reduced number of tasks needed to compute distributed $\mat{B}$ ordered in colored and numbered circles, overlaying the respective data segment $\mat{B}_{ij}$ they are responsible for. Each off-diagonal task will is also responsible for computing the symmetrical $\mat{B}_{ji}$ not overlaid in the figure. A side effect of using a reduced number of tasks is that tasks now have different execution times; off-diagonal Tasks will run much longer, and in Figure~\ref{fig:task_scheudule} we illustrate (not to scale) the task scheduling for various queue sizes. Each queue has as many CUDA streams as its size $qs$, which are ordered along the vertical axis, and the horizontal axis represents execution time. This shows how using CUDA streams helps hide latency by overlapping compute and data copy, as we can see larger queues with more streams execute in a shorter time. Finally, there is an \emph{All-reduce} communication (AR) between tasks of the same number on the different GPUs to sum the local $\mat{B}_{ij}$ or $\mat{B}_{ij}$ results and to obtain the global results. All (AR) are handled with optimized and low latency NVIDIA communication collectives Library (NCCL) base communicators. The advantage of using NCCL over MPI is discussed in detail by Awan \cite{awan2016efficient}.

In the sparse case, where the product of dense factors $\mat{U}$, $\mat{V}$, and $\mat{\Sigma}$ is a significant memory bottleneck even for an out-of-memory implementation, we avoid such computation by performing all the computation at the final stage as detailed in Section\ref{sec:OOM}, which reduces to a series of matrix-vector operation instead of a matrix-matrix operation, which is efficient both computation as well as memory wise. Although pushing the computation at the later stage would require more D2H and H2D communications, these communications are smaller in size in the sparse case while enabling the decomposition of huge sparse matrices. The algorithm is presented in Algorithm~\ref{alg:dist_compute_v}. 

\subsection{Hardware and computing environment}
Benchmark tests were performed on Chicoma, a LANL internal HPC cluster composed of 118 compute nodes, with 2 AMD EPYC 7713 Processors and 4 NVIDIA Ampere A100 GPUs each. The AMD EPYC 7713 CPUs have 64 cores peaking at 3.67 GHz and 256 GB RAM memory. Each of the four NVIDIA A100 GPUs in each node provides a theoretical double-precision arithmetic capability of approximately 19.5 teraflops with 40GB VRAM memory. The nodes are networked with HPE/Cray slingshot 10 interconnect with 100Gbit/s bandwidth. Chicoma runs Shasta 1.4 OS and SLURM Job manager. 

\section{Benchmarks and Results}

\subsection{Scaling benchmarks}

\begin{figure}[ht!]
\centering
    \begin{minipage}[b]{0.45\linewidth}
        \includegraphics[width=1\linewidth]{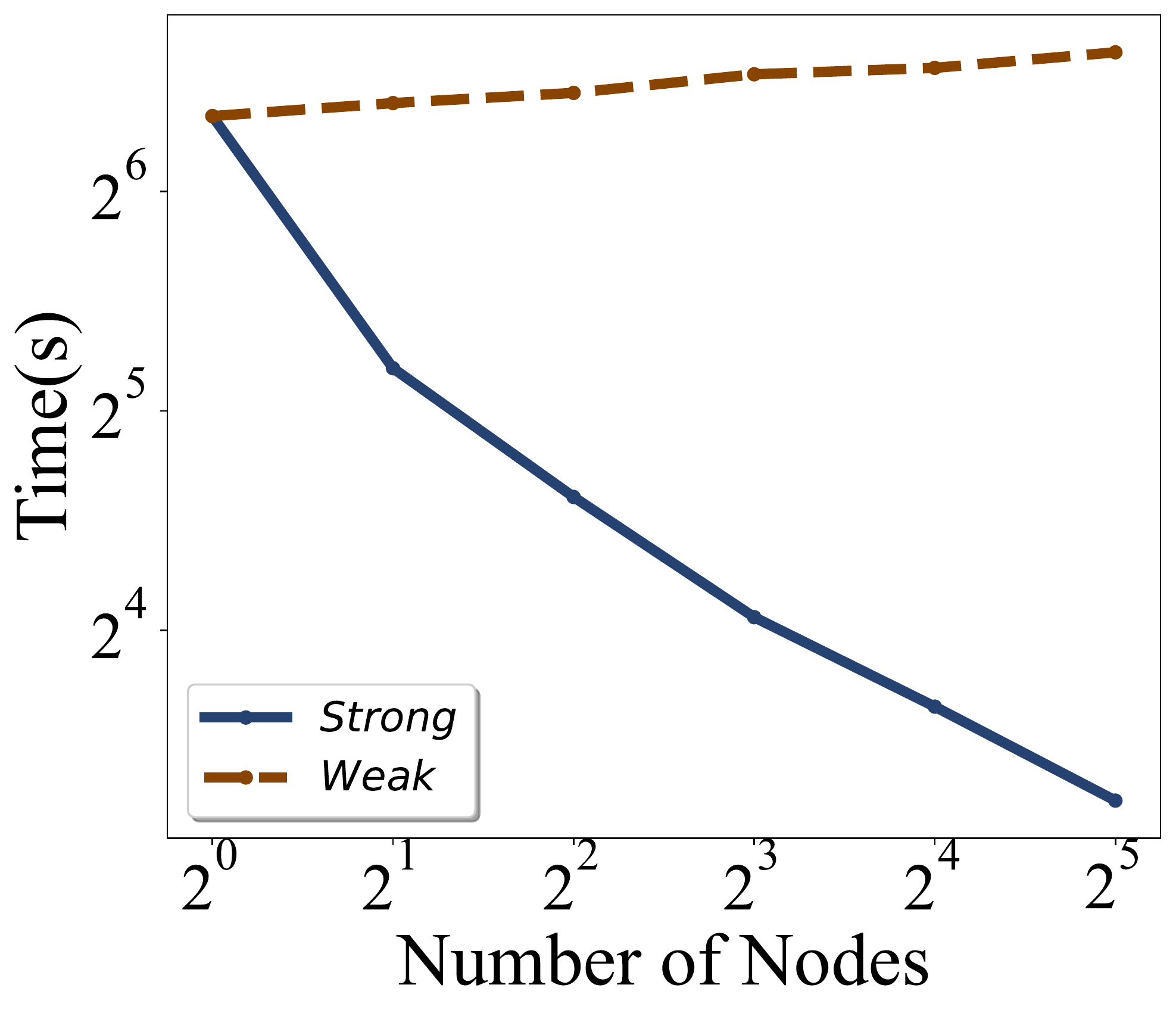}
        \subcaption{\label{fig:bench_dense}} 
    \end{minipage}
    \begin{minipage}[b]{0.45\linewidth}
        \includegraphics[width=1\linewidth]{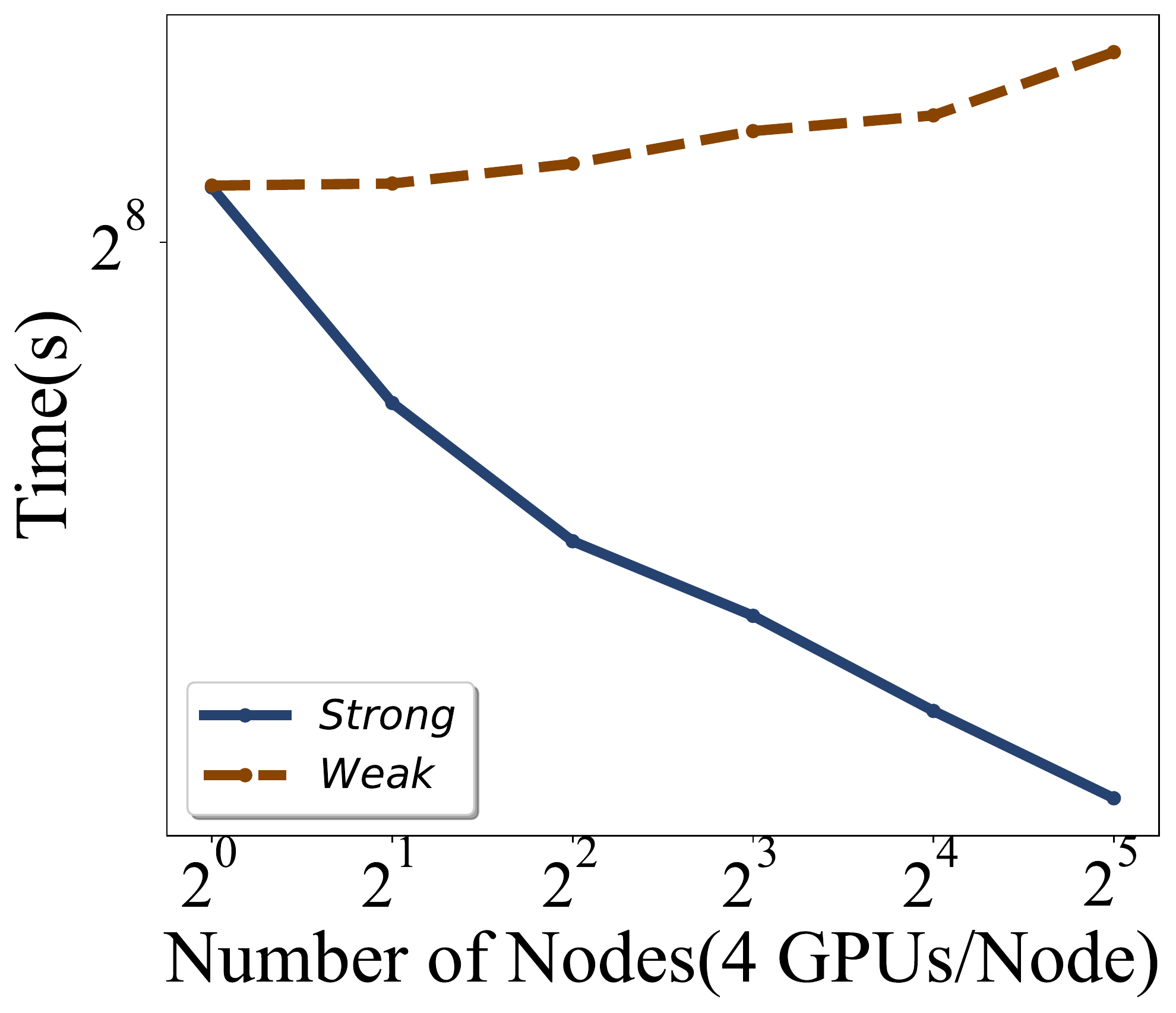}
        \subcaption{\label{fig:bench_sparse}} 
    \end{minipage}

    \caption{ SVD Scaling results for dense $\mat{A}$ in (a), and sparse $\mat{A}$ in (b). Strong scaling results shown with blue solid line, and weak scaling results with brown dashed line.} \label{fig:diributed_scaling}
\end{figure}

The first set of benchmark experiments was performed to assess the proposed implementation's strong and weak scaling characteristics on a distributed heterogeneous system. To this end, we used a dense matrix $\mat{A}$ of shape $(m,n)= [262144, 32768]$ and a sparse matrix of shape $(m,n)=[33554432,33554432]$ per node. The former had a size $S_A = 32 GB$ in single precision, and the latter a size $S_A = 4 PB$. For the strong scaling experiment in both sparse and dense datasets, the largest data that fits in the GPU memory of a node was selected. The sparse matrix was randomly generated with a density $\delta \approx 10^{-6}$ and stored in a sparse Compressed Sparse Row (CSR) format with size $S_s \approx 4GB$. For the strong scaling, the data shape per $N_n$ node is given as $(m/N_n,n)$, whereas for the weak scaling, each node has the same data shape of $(m,n)$. The scaling experiments were performed for various node counts $N_n$ = [1,2,4,8,16,32], where each node has 4 GPUs, so the minimum and the maximum number of GPUs utilized were $N=4$ and $N=128$, respectively. The largest data size for weak scaling in dense scenario is $32\times 32GB=1TB$ for 32 nodes whereas for the sparse scenario is $32\times 4PB=128~PB$ with compressed size of $32\times 4=128~GB$. The truncated SVD was evaluated for $k=32$, and both batch size and queue size were set to 1 making these implementations purely distributed without out of memory feature. Early loop termination in Algorithm~\ref{alg:svd_1d}(line 10-15) due to convergence is avoided by disabling convergence criterion at lines (13-14) of Algorithm~\ref{alg:svd_1d}. When $\mat{A}$ is dense, the distributed Gram is directly computed using Algorithm~\ref{alg:dist_gram}, and when $\mat{A}$ is sparse, Algorithm~\ref{alg:dist_compute_v} is used to avoid computing the distributed Gram directly.

Strong and weak scaling benchmark results for dense $\mat{A}$ are shown in the graph of total execution time vs. number of nodes in Figure(\ref{fig:bench_dense}). Similarly, the strong and weak scaling benchmark results for sparse $\mat{A}$ are shown in Figure(\ref{fig:bench_sparse}). 
Weak scaling results (brown dashed line) for dense and sparse $\mat{A}$ indicate good scaling maintained up to a node count of $N_n \approx 8$. The performance drop at higher node counts typically indicates latency due to communication becoming increasingly significant at higher node counts. 
Further, weak and strong scaling results obtained for dense $\mat{A}$ appear to be better than those obtained with sparse $\mat{A}$. This is to be expected as Algorithm~\ref{alg:dist_gram} computes the Gram matrix just once and reuses it in every iteration at line 11 of Algorithm~\ref{alg:svd_1d}. On the other hand, by avoiding computing $\mat{A}^T\mat{A}$  with Algorithm~\ref{alg:dist_compute_v}, communication takes place when performing the two separate All-reduce-sum operations at lines 6 and 8 of Algorithm~\ref{alg:dist_compute_v} aggregate with each iteration. Also, note that additional latency resulting from having to batch $\mat{V}$ In OOM degree 1 scenarios will affect the performance of the Algorithm~\ref{alg:dist_compute_v}.

\subsection{Out of Memory benchmarks}
\begin{figure}[ht!]
\centering
         \begin{minipage}[b]{0.45\linewidth}
        \includegraphics[width=1\linewidth]{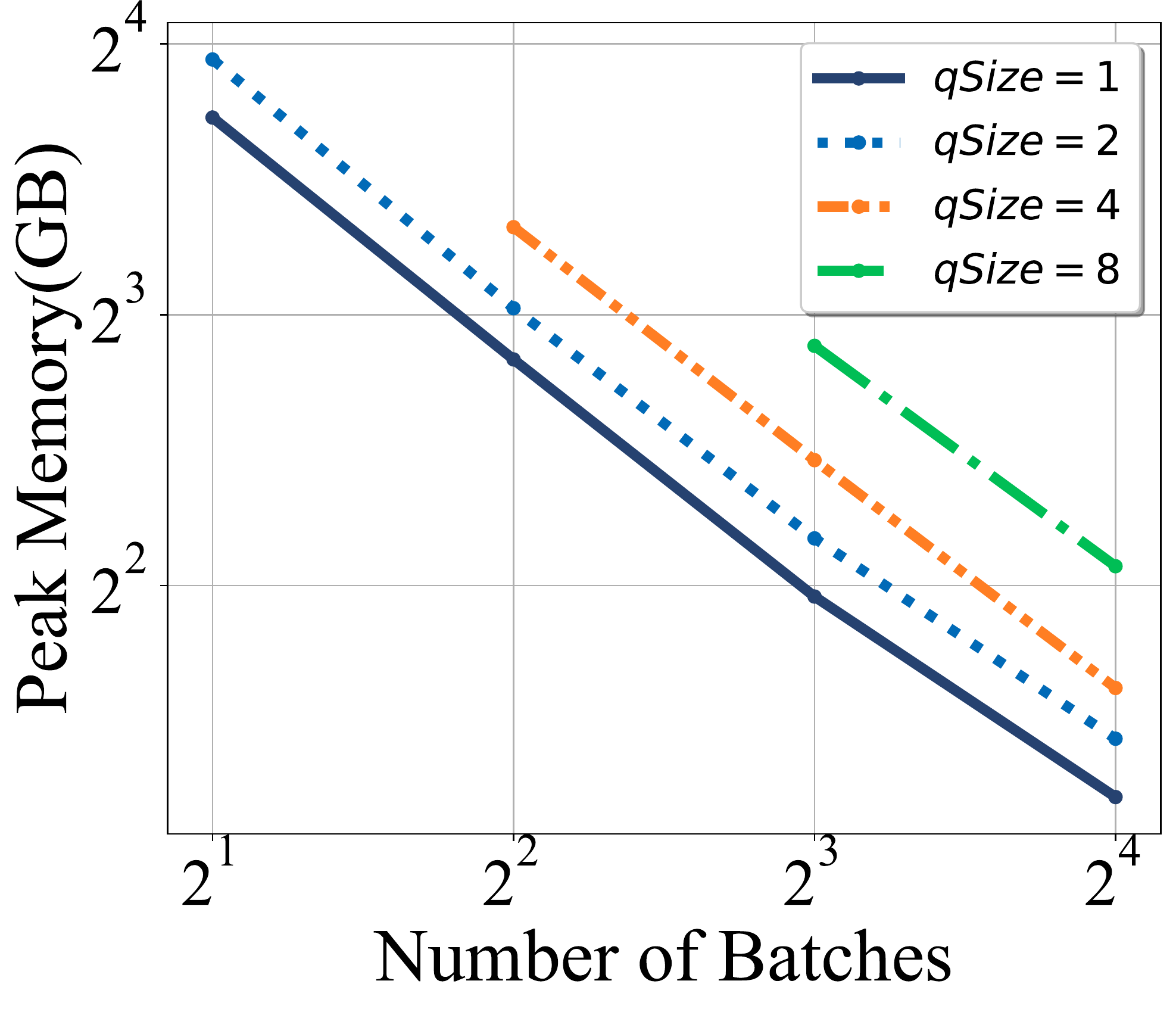}
        \subcaption{\label{fig:OOM_PeakMem_batch}} 
        \end{minipage}
    \begin{minipage}[b]{0.45\linewidth}
        \includegraphics[width=1\linewidth]{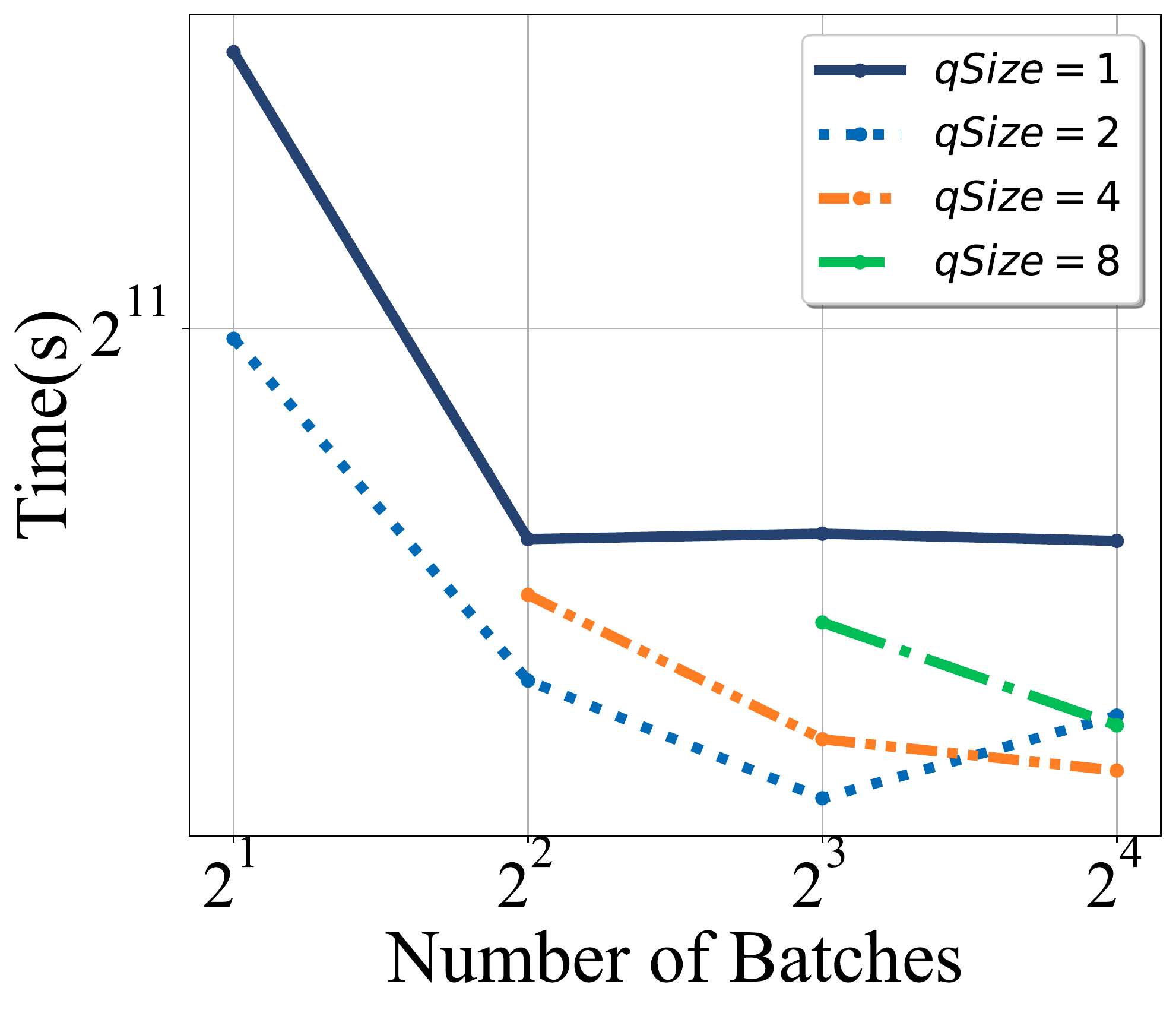}
        \subcaption{\label{fig:OOM_Peaktime_batch}} 
    \end{minipage}

    \caption{ (a) OOM SVD peak memory vs number of batches for different queue sizes and (b) OOM SVD time vs number of batches for different queue sizes . The two figures dictate the selection of the right batch size and queue size for faster decomposition, given a peak memory requirement for given data size. }  \label{fig:diributed_OOM1}
\end{figure}
Next, we assess the effectiveness of the proposed batching technique for OOM scenarios and the use of the CUDA stream queues to reduce communication in Algorithm~\ref{alg:dist_compute_v}. To this end, the proposed implementation using Algorithm~\ref{alg:dist_compute_v} is tested in an OOM scenario of degree 1, where the same sparse matrix used in the benchmarks above is decomposed up to $k=32$ with the number of iterations in Algorithm~\ref{alg:svd_1d}(line 10-15) fixed to 100. Light arrays $\mat{A}$, $\mat{U}$, $\mat{\Sigma}$ are cached on GPU memory, and heavy co-factor $\mat{V}$ is stored on the host. This means that during block-operations, $\mat{A}$ residing on GPU will be tiled, while $\mat{V}$ residing on the host is batched to GPU. For this experiment, the number of nodes is fixed to $N_n=2$ to ensure that the algorithm is distributed with inter-node communication taking place and being accounted for in performance evaluation. The execution time of the SVD algorithm and the corresponding peak memory utilization per GPU as a function of number of batches ($n_b=[2,4,8,16]$) for various queue sizes ($q_s=[1,2,4,8]$) are reported in Figure(\ref{fig:diributed_OOM1}).  

In Figure(\ref{fig:OOM_PeakMem_batch}), we show Peak memory utilization vs number of batches for various queue sizes. First, we note that results are reported such that the queue size is at most equal to the number of batches; this is to avoid unnecessary buffer reservation when $q_s > n_b$, which will skew the peak memory utilization. Second, we note a decrease in Peak memory utilization with an increasing number of batches for any fixed queue size, which is expected as increasing the number of batches results in processing smaller batches. Third, we also observe an increased peak memory utilization with increased queue size for any fixed batch number, which we can understand as the aggregated peak memory utilization caused by the different batches. The takeaway is that when an SVD is not possible with given batch size, increasing the number of batches can lower the peak memory per GPU down to levels where SVD becomes feasible. One thing to keep in mind is that ultimately all the curves in Figure(\ref{fig:OOM_PeakMem_batch}) will asymptotically decay to a minimum equal to the cumulative sum of the sizes of all arrays cached on GPU.
In Figure(\ref{fig:OOM_Peaktime_batch}) we show SVD time vs number of batches for various queue sizes. First, we see that it is, in all cases, a good idea to choose a queue size $q_s>1$ if one wants to speed up the SVD calculation. This is explained by using large stream queue sizes makes more streams available to overlap memory copies, all-reduce communications, and compute concurrently. It is, however, not the case that more streams will always make this process better, as we can see it not being the case when $n_b=8$, where the SVD time when $q_s=2$ is lower than when $q_s=4$ which in turn is lower than when $q_s=8$. This is explained by the fact that CUDA core counts are finite and that some streams will block and wait when all cores are busy processing other streams, causing load balancing delays. Consequently, it is important to adequately try to fine-tune $q_s$ and $n_b$ for optimal performance.

\section{Conclusion}
We have presented a distributed out-of-memory implementation of the truncated SVD based on the power method. The potential memory utilization hot spots of the original power method were discussed and found to occur while computing the residual or the Gram matrix. We addressed this concern analytically, redesigning the algorithm by directly updating the singular vectors, eliminating the need to compute both the residual and the gram matrices. Strong and weak scaling results were presented to demonstrate the scalability of the modified algorithm relative to the original implementation. Benchmark results were shown for the case of a sparse matrix of size 128~PB with density $10^{-6}$ compressed to a CSR format of size 128~GB, decomposed to a rank k=32. The efficacy of batching employed in the proposed implementation to manage peak memory utilization and use of CUDA streams to hide data transfer and communication latency were shown and discussed through the benchmark results.

\section{Acknowledgements}
This research was funded by Laboratory Directed Research and Development  (20190020DR), and resources were provided by the Los Alamos National Laboratory Institutional Computing Program, supported by the U.S. Department of Energy National Nuclear Security Administration under Contract No. 89233218CNA000001. The work of Hristo Djidjev has been also partially supported by Grant No.\ BG05M2OP001-1.001-0003, financed by the Science and Education for Smart Growth Operational Program (2014-2020) and co-financed by the European Union through the European structural and Investment funds and by Grant No KP-06-DB-11 of the Bulgarian National Science Fund.

\bibliographystyle{plain}
\bibliography{main}

\end{document}